\documentclass{article}
\usepackage{url}
\usepackage{graphicx}
\usepackage[lined,boxed,linesnumbered]{algorithm2e}

\usepackage{authblk}
\usepackage{setspace}
\usepackage{amsmath}

\usepackage{color}
\begin{document}
\title{Identifying a set of influential spreaders in complex networks }

\author{Jian-Xiong Zhang$^{1,2}$}
\author{Duan-Bing Chen$^{1,2\ast}$}
\author{Qiang Dong$^{1,2}$}
\author{Zhi-Dan Zhao$^{2}$}

\affil{
$^{1}$ Web Sciences Center, University of Electronic Science and Technology of China, Chengdu 611731, P.R. China\\
$^{2}$ Big Data Research Center, University of Electronic Science and Technology of China, Chengdu 611731, P.R. China\\
$^{\ast}$ Correspondence should be addressed to dbchen@uestc.edu.cn
}
\maketitle
\section*{Abstract}

Identifying a set of influential spreaders in complex networks plays a crucial role in effective information spreading. A simple strategy is to choose top-$r$ ranked nodes as spreaders according to influence ranking method such as PageRank, ClusterRank and $k$-shell decomposition. Besides, some heuristic methods such as hill-climbing, SPIN, degree discount and independent set based are also proposed. However, these approaches suffer from a possibility that some spreaders are so close together that they overlap sphere of influence or time consuming. In this report, we present a simply yet effectively iterative method named VoteRank to identify a set of decentralized spreaders with the best spreading ability. In this approach, all nodes vote in a spreader in each turn, and the voting ability of neighbors of elected spreader will be decreased in subsequent turn. Experimental results on four real networks show that under Susceptible-Infected-Recovered (SIR) \textcolor[rgb]{1.00,0.00,0.00}{ and Susceptible-Infected (SI) models}, VoteRank outperforms the traditional benchmark methods on both \textcolor[rgb]{1.00,0.00,0.00}{spreading rate} and final affected scale. What's more, VoteRank \textcolor[rgb]{1.00,0.00,0.00}{has superior computational efficiency.}

\section*{Introduction}
\textcolor[rgb]{1.00,0.00,0.00}{In real world, many complex systems can be represented as complex networks\cite{Boccalettia2006PhysRep,Gao2015SciRep,Gao2015ExpTher,Gao2012NonAna,pastor2001epidemic,Luo2015PLoSONE}, in which,} Many activities such as advertising over media and word-of-mouth on social networks can be described by information spreading on complex networks\cite{pastor2001epidemic,LvNJP2011,MyersKDD2012,LiuZhang2014,Cinimi2012PRE,Chen2014SciRep}. Maximizing the scale of spreading is a common \textcolor[rgb]{1.00,0.00,0.00}{target}. If a market manager want to advertise a new product on Twitter.com, \textcolor[rgb]{1.00,0.00,0.00}{she/he tries} to choose a small number of users to provide them \textcolor[rgb]{1.00,0.00,0.00}{with free products in exchange for posting  tweets} about the product to influence their friends to buy the products. So, the task of market manager is to choose \textcolor[rgb]{1.00,0.00,0.00}{a few users such that the product information can be transmitted to more users and, more products can be sold finally}. With the topology unchanged or changed slightly, the location of  source spreaders determines the final scale of spreading on large degree. The problem of choosing initial nodes as source spreaders to achieve maximum scale of spreading is defined as  \emph{influence~maximization~problem} \cite{kempe2003maximizing}. Our research focuses on the strategy of choosing a set of critical  nodes as source spreaders in this report.

As influential nodes have \textcolor[rgb]{1.00,0.00,0.00}{strong} ability to \textcolor[rgb]{1.00,0.00,0.00}{affect} other nodes, selecting top-ranked influential nodes as source spreaders is a common and \textcolor[rgb]{1.00,0.00,0.00}{classical} strategy. Up to now, \textcolor[rgb]{1.00,0.00,0.00}{many ranking methods }have been proposed, such as degree, closeness\cite{sabidussi1966centrality},  betweenness\cite{Freeman1979} centralities, and other heuristic algorithms\cite{EPL2013Chen, EPL2014Ren, Chen2012PhysicaA, AskariSichani2015}. Random-walk based methods such as well-known PageRank\cite{PageRank1999} and LeaderRank\cite{lu2011leaders} \textcolor[rgb]{1.00,0.00,0.00}{have been receiving great attentions and shown} significant value in last few years. Pei et al. \cite{PeiSR2014} addressed a direct method to search for influential spreaders by following the real spreading dynamics in a wide range of networks.  Some other methods such as HITS\cite{kleinberg1999authoritative} and TwitterRank\cite{weng2010twitterrank} are also useful and effective. Recently a local based method ClusterRank\cite{chen2013identifying} has also good performance in some cases. Ref.\cite{kitsak2010identification} shows that the crucial factor of node's influence is its location in network measured by $k$-shell value. Under this measuring strategy, nodes with larger $k$-shell values usually have more ability on spreading. Wei et al. \cite{Wei2015PhysicaA} proposed a weighted $k$-shell decomposition to identify influential nodes. Liu et al. \cite{Liu2015SR} introduced a measure based on link diversity of shells to distinguish the true core and core-like group so as to find the real influential spreaders. Based on Ref. \cite{Liu2015SR}, Liu et al. \cite{Liu2015SR_b} proposed an improved $k$-shell method by removing redundant links. L\"{u} et al. \cite{Lv2016NC} unveiled the elegant mathematical relationship among three simple yet important centrality measures of networks, i.e., degree, H-index and coreness. Ref. \cite{kitsak2010identification} indicates that top-ranked nodes obtained by $k$-shell decomposition are of significant influences. However, if select them as a group to spread, the result is not so good, even is worse than the result of pure degree centrality. Like $k$-shell method, other ranking methods such as closeness, PageRank, LeaderRank and ClusterRank suffer the similar limitation.

Kempe et al. \cite{kempe2003maximizing} proposed a hill-climbing based greedy algorithm that can find a group of important nodes to affect the widest scope of nodes. Their work can overcome the shortcoming mentioned before. However, it is very time consuming, especial in large scale networks. Based on greedy strategy, Narayanam and Narahari \cite{SPIN2011} proposed a much faster algorithm SPIN approach than greedy algorithm while its quality decreasing a little. Unfortunately, SPIN is also hardly applied to large scale networks. For example, the CPU running time is 28.25 minutes if to find top-30 important nodes in network with 1589 nodes. For this reason, some fast heuristic algorithms are presented in recent years. Chen et al. \cite{Chen2009KDD} proposed degree discount heuristic algorithm, which nearly matches the performance of the greedy methods for the IC model. Tang et al. \cite{TangSIGMOD2014} presented a Two-phase Influence Maximization (TIM) algorithm that aimed to bridge the theory and practice in influence maximization. In theory, TIM runs in $O((r+\ell)(n+m)\log n/\varepsilon^2)$ expected time and returns a $(1-1/e-\varepsilon)$-approximate solution with at least $1-n^{-\ell}$ probability where $\ell$ and $\varepsilon$ are parameters. Zhao et al. \cite{ZhaoEPL2014} made an attempt to find a set of important spreaders by generalizing the idea of the coloring problem in graph theory \textcolor[rgb]{1.00,0.00,0.00}{\cite{Welsh1967Comput}} to complex networks. Ji et al. \cite{Huarxiv2015} proposed an effective multiple leaders identifying method based on percolation theory. The method well utilizes the similarities between the pre-percolated state and the average of information propagation in each social cluster to obtain a set of distributed and coordinated spreaders. \textcolor[rgb]{1.00,0.00,0.00}{Very recently, Morone and Makse presented an effective method to find a set of critical nodes by mapping the problem onto optimal
percolation in random networks\cite{Morone2015Nature}.} He et al. \cite{HePLoS2015} proposed a novel method to identify multiple spreaders in complex networks with community structures.

In this report, we propose a simply yet effectively iterative method named VoteRank to choose a set of influential spreaders. In our method, influential spreaders are elected one by one according to their voting scores obtained from their neighbors. At each iteration, the voting ability of elected spreader will be set to zero while that of \textcolor[rgb]{1.00,0.00,0.00}{its} neighbors will be decreased by a factor. Our method can be applied to large scale network with millions of nodes since it just updates local information after selecting a spreader. Experimental results on real datasets show that our method outperforms traditional methods \textcolor[rgb]{1.00,0.00,0.00}{on both} final affected scale and  \textcolor[rgb]{1.00,0.00,0.00}{spreading rate}. What's more, VoteRank is also superior to other group-spreader identifying methods on computational time.

\section*{Methods and Materials}
\subsection*{Spreading Models}

In this report, we \textcolor[rgb]{1.00,0.00,0.00}{mainly} use SIR epidemic model \textcolor[rgb]{1.00,0.00,0.00}{with limited contact} \cite{tao2006epidemic,hethcote2000mathematics} to  evaluate methods. In SIR model, each node is in one of three statuses, i.e., Susceptible(S), Infected(I) and Recovered(R). Initially, all nodes are  susceptible status except for a set of $r$ infected nodes selected as source spreaders. At each time step, infected node tries to infect one of its neighbors with probability $\mu$. At the same time, each infected node will be recovered with a probability $\beta$, if success, it won't be infected again and no longer infect other susceptible nodes. The process terminates if there isn't any infected node in network. In this report, we use $\lambda=\mu/\beta$ to represent \emph{infected rate}, which is crucial to infected speed and final affected scale that are often used to indicate the spreading ability of $r$ source spreaders. 
\textcolor[rgb]{1.00,0.00,0.00}{Besides SIR model with limited contact, the performance of methods can also be evaluated by SIR model with full contact and SI model\cite{Barabasi1999Science}  that is usually used to evaluate the method on spreading rate especially in the early stage.}

\subsection*{VoteRank Algorithm}

 In real world, if a person A has supported person B, the support strength of A to others will fade generally. Under this perspective, a vote based approach for identifying influential spreaders named VoteRank is presented in this report. In VoteRank, the main idea is to choose a set of spreaders one by one according to voting scores of nodes obtained from their neighbors.  If we need to select top-$r$ influential spreaders, every node has to vote $r$ turns. The node getting the most votes in each turn is regarded as the most influential node in that turn and will be elected as one of top-$r$ influential spreaders. If a node has been elected as a spreader, it doesn't participate in subsequent voting, and the voting ability of its neighbors also be decreased.%
 \textcolor[rgb]{1.00,0.00,0.00}{Actually, when a node $u$ is elected as spreader, the propagation range has increased a little if the nodes near $u$ are elected as spreaders again since $u$ can transfer information to these nodes. So, it's better to select far apart nodes because they can affect as many nodes as possible. That is to say, after a node is elected as spreader, the selection probability of its neighbors and neighbors' neighbors will decrease. Under this mechanism, the selected nodes are far apart and are important in its local structure.  In fact, similar idea has been reported in references. For example, Kitsak et al. \cite{kitsak2010identification} pointed out that the propagation range would be improved greatly if any two selected spreaders are disconnected comparing with simply selecting nodes with maximum degree or $k$-shell value one by one.}

  In VoteRank, each node $u$ is attached with a tuple $(s_u, va_u)$, where voting score $s_u$ denotes the number of votes obtained from $u'$s neighbors and voting ability ${va}_u$ represents the number of votes that $u$ can give its neighbors. The details of VoteRank are described as following five steps:

step 1: Initialize. Tuples of all nodes are set to $(0,1)$.

step 2: Vote. Nodes vote for their neighbors, at the same time are voted by their neighbors. After voting step, the voting score of each node will be calculated. It is noted that the voting score of node is set to zero if it has been elected in earlier turn so as to avoid electing it again. For example, node $v_{0}$ has three neighbors $v_{1},v_{2}$ and $v_{3}$. Node $v_{0}$ will vote for $v_{1},v_{2}, v_{3}$ with $va_{v_0}$ votes, and $v_{1},v_{2}, v_{3}$ will vote for their corresponding neighbors with $va_{v_1}, va_{v_2}$ and $va_{v_3}$ votes respectively. So, the voting score of node $v_0$ is $s_{v_0} = va_{v_1}+ va_{v_2}+ va_{v_3}$. This voting process is different from political voting because some nodes just vote for less one vote in VoteRank.

step 3: Select. According to voting scores calculated in step 2, select the node $v_{max}$ that gets the most votes. This node will not participate in subsequent voting turns, that is, its voting ability $va_{v_{max}}$ will be zero from now on.

step 4: Update. Weaken the voting ability of nodes those voted for $v_{max}$ in step 2. For example, if node $u$ voted for $v_{max}$ , update the voting ability of $u$ with $va_u - f$ unless $va_u $ has been decreased to zero, where $f$ is a decreasing factor being between 0 and 1. \textcolor[rgb]{1.00,0.00,0.00}{For special case of $f = 0$, just the degree of newly elected node's neighbors will minus one since only the voting ability of newly elected node turns to zero.} In this report, \textcolor[rgb]{1.00,0.00,0.00}{we mainly focus $f$ on a simple form $\frac{1}{<k>}$,} where $<k>$ is the average degree of the network.

step 5: Repeat steps 2 to 4 until $r$ spreaders are elected.

\begin{figure}[htbp]
\centering
\includegraphics[width=12cm]{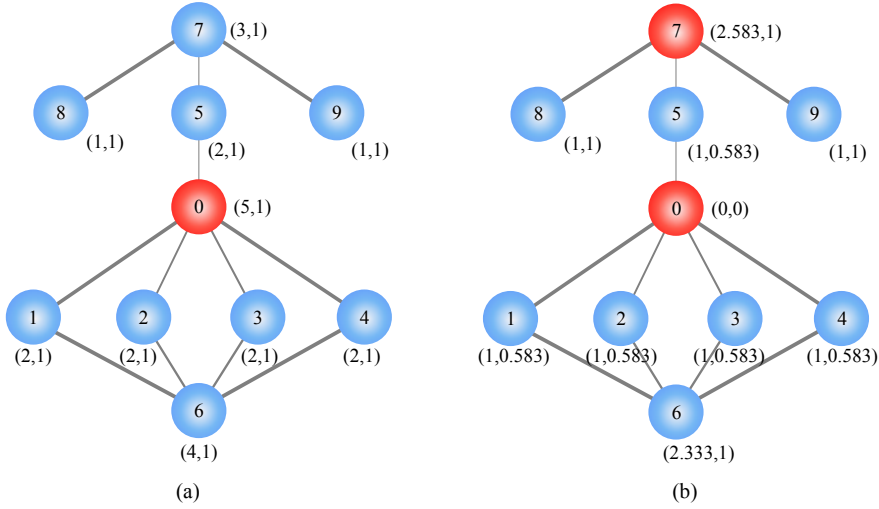}
\caption{(Color online) A toy network. In (a) node 0 is selected as one of top-2 spreaders, and in (b),  node 7 is selected as one of top-2 spreaders.}
\label{fig1:smallNetwork}
\end{figure}

In order to give an intuitive explanation, we use VoteRank to choose top-2 nodes on a small toy network with 10 nodes, as shown in Fig. \ref{fig1:smallNetwork}. Fig. \ref{fig1:smallNetwork}(a) represents the first turn of voting. The value of voting score and voting ability for each node is marked as tuple $(s, va)$ in Fig. \ref{fig1:smallNetwork}(a). In this turn, node 0 is chosen and its voting ability is set to 0. The voting abilities of nodes 1, 2, 3, 4 and 5 are reduced by $\frac{1}{2.4}=0.417$. The updated voting ability of each node is marked in Fig. \ref{fig1:smallNetwork}(b). According to new voting abilities,  node 7 is chosen since it gets the highest voting score 2.583 at the second voting.

VoteRank algorithm not only can be used to choose top-$r$ spreaders in undirected networks, but also can be used in directed networks. In directed network, if there is a link from node $u$ to node $v$, $u$ is the in-neighbor of $v$, and correspondingly, $v$ is the out-neighbor of $u$. In this report, a link from node $u$ to $v$ indicates that $v$ receives information from $u$. The directed version of VoteRank is slightly different from undirected one. \textcolor[rgb]{1.00,0.00,0.00}{Firstly,} nodes only vote for their in-neighbors, \textcolor[rgb]{1.00,0.00,0.00}{and secondly, }only the voting ability of elected node and its out-neighbors will be updated.

\subsection*{Performance Metrics}
In this report, three metrics are used to evaluate the performance of methods. The first two are based on spreading scale under SIR or SI spreading model, and the third is based on structural properties of elected spreaders.

In order to compare the spread speed for different methods, we introduce infected scale $F(t)$ at time $t$ which is defined as:
\begin{equation}\label{eq_Ft}
  F(t)=\frac{n_{I(t)}+n_{R(t)}}{n},
\end{equation}
where $n$ is the number of nodes of network, $n_{I(t)}$ and $n_{R(t)}$ \textcolor[rgb]{1.00,0.00,0.00}{($n_{R(t)=0}$ for SI model)} are the number of infected and recovered nodes at time $t$ respectively.

In order to investigate the final scale of affected nodes, final affected scale $F(t_c)$ is introduced:
\begin{equation}\label{eq_Ftc}
  F(t_c)=\frac{n_{R(t_c)}}{n},
\end{equation}
where $n_{R(t_c)}$ is the number of recovered nodes when spread process achieving steady state.

Besides $F(t)$ and $F(t_c)$, the structural properties among selected spreaders are also used to evaluate the performance of different methods. In this report, the \emph{average shortest path length} $L_s$ between each pair of source spreaders $S$ is used as evaluating metric. It is defined as:

\begin{equation}\label{eq_Ls}
  L_s =\frac{1}{|S|(|S|-1)} \sum_{\substack{u,v\in S \\ u\neq v}}l_{u,v},
\end{equation}
where $l_{u,v}$ denotes the length of the shortest path from node $u$ to $v$.

\subsection*{Data Description}

Four real networks are used to test the performance of VoteRank in this report. Networks YOUTUBE \cite{Youtube2012} and COND-MAT \cite{Cond-mat2001} are undirected and Networks BERKSTAN \cite{Berkstan2009} and NOTRE DAME \cite{Notre1999} are directed. YOUTUBE is a video-sharing web site that includes a social network, in which, nodes represent users and edges represent friendships between two users. COND-MAT is a collaboration network, which generates from the e-print arXiv and covers scientific collaborations between authors who submit papers to Condense Matter category. In BERKSTAN, nodes represent pages from berkely.edu or stanford.edu domains and directed edges represent hyperlinks between them. In NOTRE DAME network, nodes represent pages from University of Notre Dame and directed edges represent hyperlinks between them. Some topological features of these four networks, including the number of nodes $n$, the number of edges $m$, the average degree (or average out-degree for directed networks) $<k>$, the maximum degree (or maximum out-degree for directed networks) $k_{max}$, the average clustering coefficient $<c>$, and the degree heterogeneity $H$ which is defined as $\frac{<k^2>}{<k>^2}$ \textcolor[rgb]{1.00,0.00,0.00}{\cite{Hu2008PhysicaA}}, are shown in Table \ref{tb_data}.
\begin{table}
  \centering
   \caption{The basic topological features of four real networks. $n$ and $m$ are the total number of nodes and edges, respectively. $<k>$ is the average degree for undirected networks or the average out-degree for directed networks. $k_{max}$ is the maximum degree for undirected networks or the maximum out-degree for directed networks. $<c>$ is the average clustering coefficient and $H$ is the degree heterogeneity, defined as $H=\frac{<k^2>}{<k>^2}$\textcolor[rgb]{1.00,0.00,0.00}{\cite{Hu2008PhysicaA}}.  }\label{tb_data}
  \begin{tabular}{l llll lll}
  \hline
  Networks&$n$&$m$&$<k>$&$k_{max}$&$<c>$&$H$ \\ \hline
  YOUTUBE \textcolor[rgb]{1.00,0.00,0.00}{\cite{Youtube2012}} & 1134890 & 2987624 & 5.2650 & 28754 & 0.0808 & 93.9270 \\
  COND-MAT\textcolor[rgb]{1.00,0.00,0.00}{\cite{Cond-mat2001}} & 23133 & 93497 & 8.0834 & 279 & 0.6334 & 2.7305 \\
  BERKSTAN\textcolor[rgb]{1.00,0.00,0.00}{\cite{Berkstan2009}} & 685230 & 7600595 & 11.0920 & 249 & 0.5967 & 3.1744 \\
   NOTRE DAME\textcolor[rgb]{1.00,0.00,0.00}{\cite{Notre1999}} & 325729 & 1497134 & 4.5120 & 3444 & 0.2346 & 23.4647 \\
  \hline
\end{tabular}

\end{table}


\section*{Results}

\begin{figure}[htbp]
\centering
\includegraphics[width=12cm]{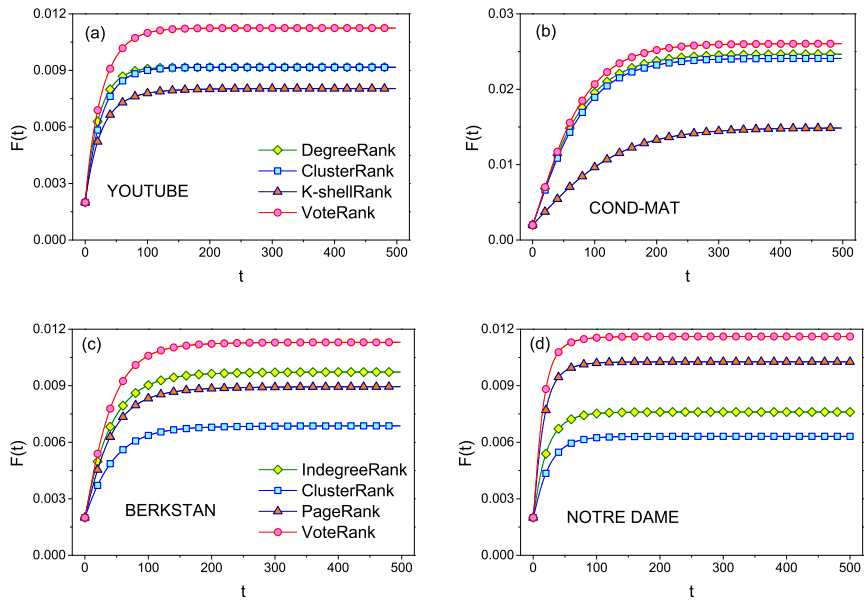}
\caption{(Color online) The infected scale $F(t)$ on four networks under different methods, where $\lambda=1.5$ and $p=0.002$. \textcolor[rgb]{1.00,0.00,0.00}{The results are averaged over 100 independent runs.}}\label{fig2:spreadscale}
\end{figure}

The performances of VoteRank and other methods are evaluated by three metrics mentioned before on four real networks. Figure \ref{fig2:spreadscale} shows the infected scale $F(t)$ on four networks under different methods with infected rate $\lambda=1.5$ and $p=0.002$ where $p$ is the ratio of the number of source spreaders and that of nodes in network. From Fig. \ref{fig2:spreadscale}, it can be seen that from the source spreaders obtained by VoteRank, information can spread faster and eventually affect larger scale than that by other methods. \textcolor[rgb]{1.00,0.00,0.00}{Moreover, the deviation of $F(t)$ is generally small especial for YOUTUBE, BERKSTAN and NOTRE DAME. }

\begin{figure}[htbp]
\centering
\includegraphics[width=12cm]{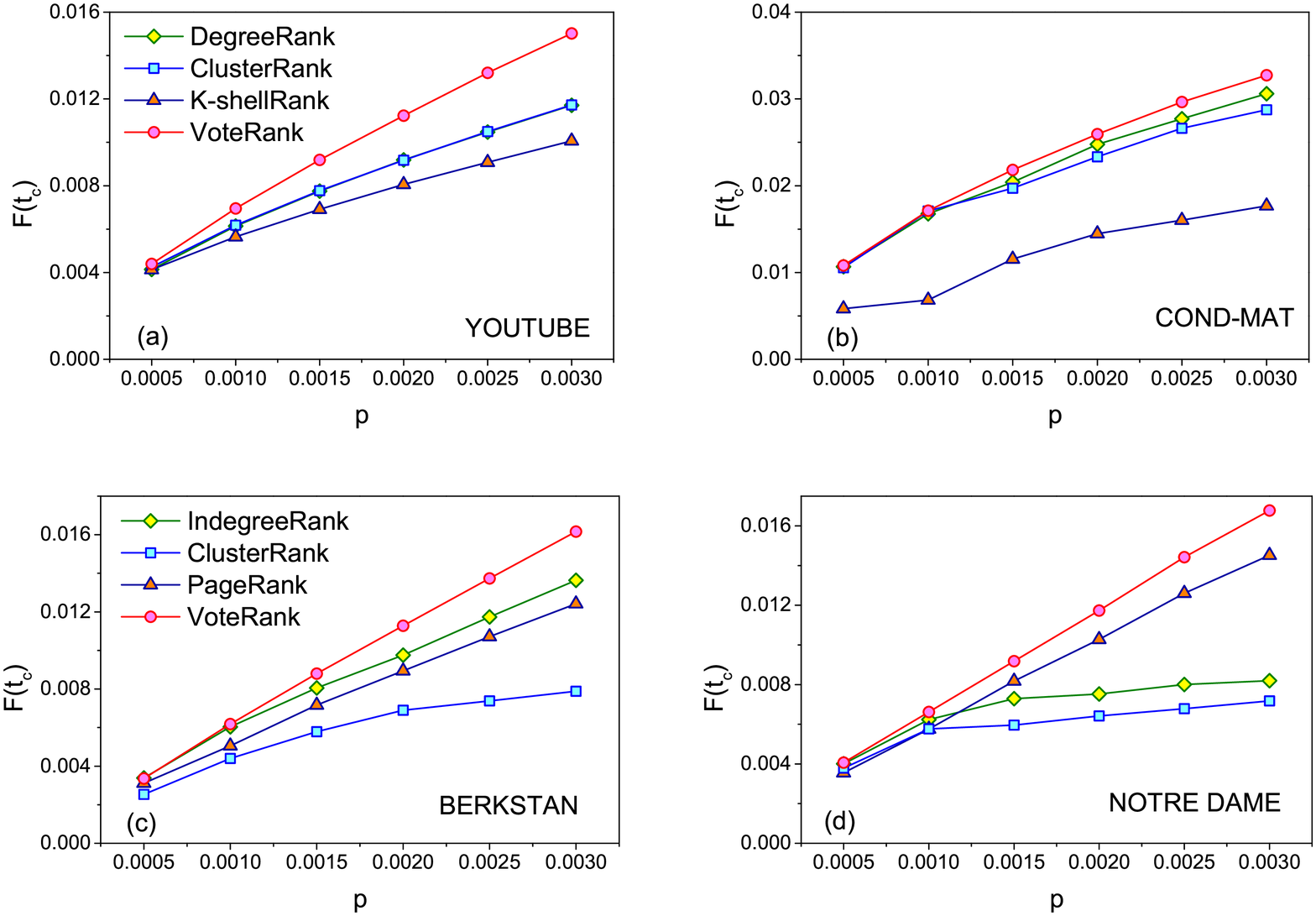}
\caption{(Color online) The final affected scale $F(t_c)$ with different number of source spreaders. In (a)-(d), $\lambda=1.5$, in (a) and (b),  $\beta=\frac{1}{<k>}$ and in (c) and (d), $\beta=\frac{1}{<k^{out}>}$. \textcolor[rgb]{1.00,0.00,0.00}{The results are averaged over 100 independent runs.}}\label{fig3:finalspread}
\end{figure}

Figure \ref{fig3:finalspread} shows the final affected scale $F(t_c)$ with different number of source spreaders. It's obvious that VoteRank can achieve wider final affected scale $F(t_c)$ than other benchmark methods under same number of source spreaders, especially when the number of source spreaders is large.

\begin{figure}[htbp]
\centering
\includegraphics[width=12cm]{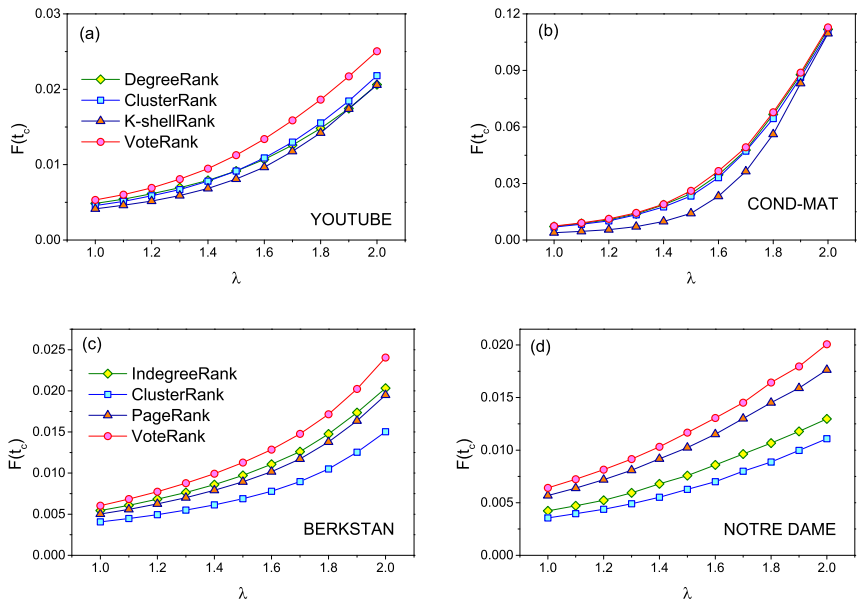}
\caption{(Color online) The final affected scale $F(t_c)$ with different infected rate $\lambda$ where $p=0.002$. \textcolor[rgb]{1.00,0.00,0.00}{The results are averaged over 100 independent runs.}}\label{fig4:spreadLambda}
\end{figure}

Figure \ref{fig4:spreadLambda} shows the $F(t_{c})$ with different $\lambda$ for different methods on four networks. From Fig. \ref{fig4:spreadLambda}, it can be seen that VoteRank can achieve wider spread scale than other methods under different $\lambda$, especial in YOUTUBE, BERKSTAN and NOTRE DAME networks. If $\lambda$ is too small, information can not be effectively spread no matter how to choose source spreaders. And if $\lambda$ is too large, information can spread all over the network. For this reason, $\lambda$ just be ranging from 1 to 2 in this report so as to compare the difference of methods clearly.

Actually, final affected scale $F(t_{c})$ is not only determined by the influence of source spreaders, but also by their relative location. For this reason, $k$-shell decomposition can dig out influential single spreader effectively, but perform poorly on selecting group spreaders by simply selecting nodes with the biggest $k$-shell value. \textcolor[rgb]{1.00,0.00,0.00}{To overcome this limitation in some degree, a reasonably improved strategy is to choose nodes with the highest voting score or $k$-shell value such that any two of selected spreaders are not directly linked. That is, under current state, if a node with the highest score is the neighbor of any selected spreader, we will skip this node and consider the next one. Under this improved strategy, VoteRank and k-shellRank can be modified as their improved versions, i.e., VoteRank-Non and K-shellRank-Non, respectively.} In order to evaluate the performance of VoteRank with this improved selecting strategy, we compare K-shellRank and VoteRank with K-shellRank-Non and VoteRank-Non.

\begin{figure}[htbp]
\centering
\includegraphics[width=12cm]{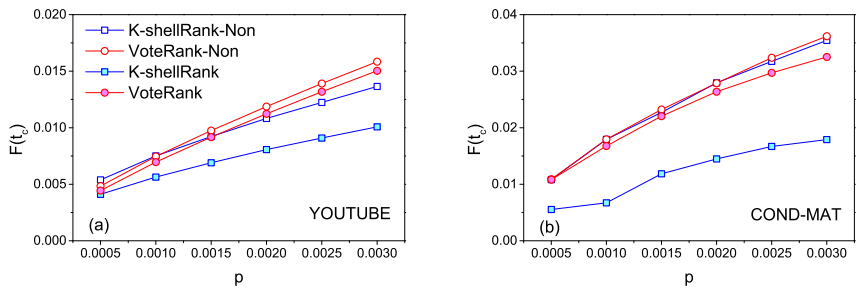}
\caption{(Color online) The final affected scale $F(t_c)$ under different initial infected scale $p$. Both in (a) and (b), $\lambda=1.5$, and K-shellRank-Non and VoteRank-Non are improved versions of K-shellRank and VoteRank, in which, any two spreaders are not \textcolor[rgb]{1.00,0.00,0.00}{directly linked. The results are averaged over 100 independent runs.}}\label{fig5:SpreadInitScale}
\end{figure}

Figure \ref{fig5:SpreadInitScale} shows the results of $F(t_c)$ against $p$ ranging from 0.0005 to 0.003 on two undirected networks. Both K-shellRank-Non and VoteRank-Non are improved compared with K-shellRank and VoteRank. Particularly, K-shellRank gets significant improvement. Even though, VoteRank-Non outperforms K-shellRank-Non when the number of source spreaders is large, especial in YOUTUBE. The results of VoteRank-Non and original VoteRank are very close, as shown in Fig. \ref{fig5:SpreadInitScale}. This indicates that the source spreaders selected by VoteRank are more disperse and diverse than K-shellRank. Interestingly, VoteRank even outperforms K-shellRank-Non when $p$ is larger than 0.0015 in YOUTUBE, as shown in Fig. \ref{fig5:SpreadInitScale}(a).

\begin{figure}[htbp]
\centering
\includegraphics[width=12cm]{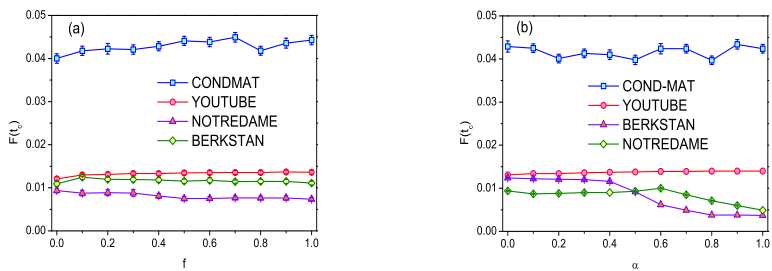}
\caption{\textcolor[rgb]{1.00,0.00,0.00}{(Color online) The final affected scale $F(t_c)$ with (a) different decreasing factor $f$ and (b) different initial voting ability where $\lambda=1.5$ and $p=0.002$. The results are averaged over 100 independent runs.}}\label{fig6:effectpara}
\end{figure}

\textcolor[rgb]{1.00,0.00,0.00}{In VoteRank, there are two parameters, i.e., decreasing factor $f$ and initial voting ability. The final affected scale $F(t_c)$ under different $f$ is compared, as shown in Fig. \ref{fig6:effectpara}(a). From this figure, it can be seen that the final affected scale $F(t_c)$ for $f>0$ is larger than that for $f=0$ except for NOTREDAME. The effect of initial voting ability on VoteRank is also analyzed. The initial voting ability of node $i$ is set as $k_i^{\alpha}$  ($(k_i^{out})^{\alpha}$  for directed network) where $\alpha$  is a parameter whose value is from zero to one, correspondingly. For curtain $\alpha$,  the parameter $f$ of node $i$ is set as $\frac{k_i^{\alpha}}{<k>}$ ( $\frac{(k_i^{out})^{\alpha}}{<k^{out}>}$ for directed network). The initial voting ability is 1 when $\alpha=0$ , and it equals node degree when  $\alpha=1$. The effect of initial voting ability is shown in Fig. \ref{fig6:effectpara}(b). Generally, in undirected networks, initial voting ability has little effect on $F(t_c)$. However, in directed networks, the smaller initial voting ability is a relatively better choice.}

\begin{figure}[htbp]
\centering
\includegraphics[width=12cm]{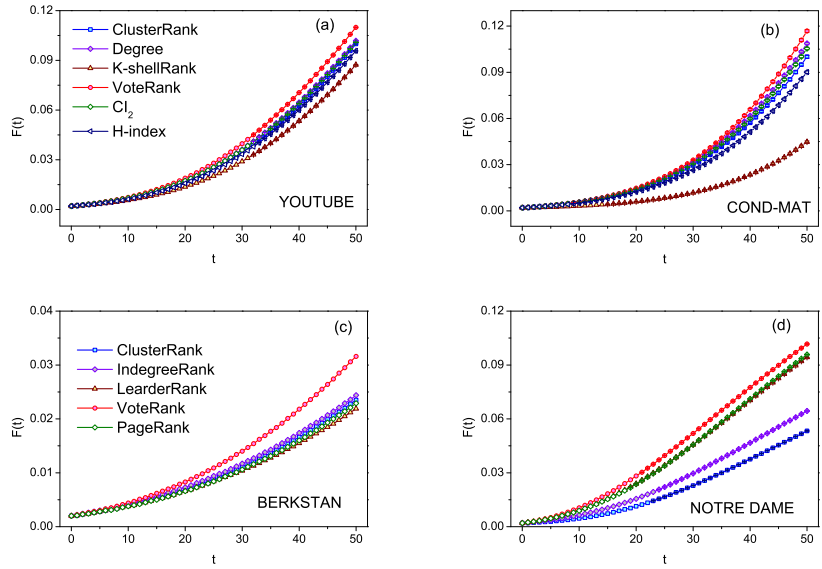}
\caption{\textcolor[rgb]{1.00,0.00,0.00}{(Color online)  The infected scale $F(t)$ on SI model with $\lambda=1.5$  , $p = 0.002$, and $\beta=\frac{1}{<k>}$  in (a) and (b),   $\beta=\frac{1}{<k^{out}>}$ in (c) and (d). The results are averaged over 100 independent runs.}}\label{fig7:SImodel}
\end{figure}

\textcolor[rgb]{1.00,0.00,0.00}{Besides SIR model with limited contact, the performance of methods are also compared on other spreading models such as SI model and SIR model with full contact process, in which, a node will contact its all neighbors. SI model is usually used to evaluate the method on spreading rate especially in the early stage. In SI model, the infected scale $F(t)$ of early stage of different methods is compared, as shown in Fig. \ref{fig7:SImodel}. From this figure, it can be seen that from the source spreaders obtained by VoteRank, the information will spread faster than that from other methods. The performance of VoteRank on SIR spreading model with full contact process with $\beta=1$, $\lambda=1.5\lambda_c$ is compared with other methods where $\lambda_c$ is the threshold \cite{Castellano2010PRL,Chu2015JBios,Li2013PRE}. The results of final affected scale $F(t_c)$ for different methods are shown in Table \ref{tb_SIRFull}. From this table, it can be seen that in most of cases, VoteRank is rather good. }

\begin{table}
  \centering
  \textcolor[rgb]{1.00,0.00,0.00}{
   \caption{The final affected scale $F(t_c)$ for different methods on SIR spreading model with full contact process where $\lambda=1.5\lambda_c$, $\beta=1$  and $p = 0.003$ . The results are averaged over 100 independent runs. }\label{tb_SIRFull}
  \begin{tabular}{l llll }
  \hline
 &         YOUTUBE&	COND-MAT &	BERKSTAN&	NOTRE DAME\\ \hline
Degree                                      &	0.0065 &	0.1213&	0.2117&	0.0365\\
ClusterRank\cite{chen2013identifying}	    &0.0055	&0.1181	&0.2018	&0.0326\\
KshellRank \cite{kitsak2010identification}	&0.0054&	0.1198	& /$^{\dag}$  &	/    \\
CI$_2$ \cite{Morone2015Nature}              &	0.0051 &	0.1183		& / &  / \\
H-index \cite{Lv2016NC}	                    &0.0045	&0.1196	& / & /	\\
PageRank \cite{PageRank1999}                &	/&  / &		0.2045&	0.0415\\
LeaderRank\cite{lu2011leaders}	            &	/	& / & 0.2037	& 0.0423 \\
VoteRank	                                &0.0064&	0.1239&	0.2190&	0.0385 \\
  \hline
\end{tabular}}

\textcolor[rgb]{1.00,0.00,0.00}{$^{\dag}$ Just undirected versions of KshellRank, CI$_2$ and H-index and directed versions of PageRank and LeaderRank are considered in this report.}
\end{table}

\begin{figure}[htbp]
\centering
\includegraphics[width=12cm]{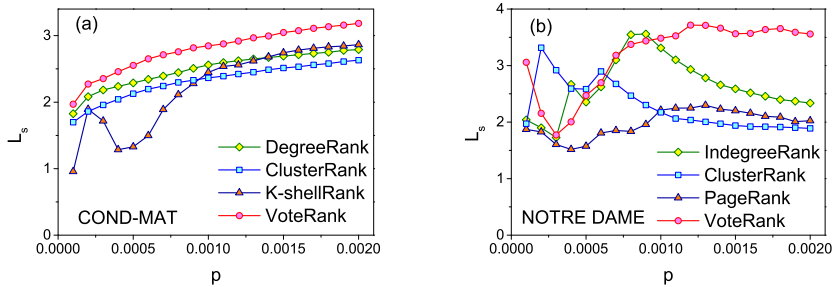}
\caption{(Color online) Average shortest path length $L_s$ for different methods under different scale of source spreaders.}\label{fig6:seedsLength}
\end{figure}

To verify  source spreaders selected by VoteRank are more scattered than that by other methods, the \emph{average shortest path length} $L_s$  obtained by VoteRank and other methods are compared. We just use two small networks, NOTRE DAME(directed) and COND-MAT(undirected) to analyze in this report for calculating the length of the shortest path in large scale network being time consuming. Figure \ref{fig6:seedsLength} shows $L_s$ of source spreaders selected by different methods under different scale of source spreaders. From Fig. \ref{fig6:seedsLength}, it can be seen that spreaders selected by VoteRank have larger $L_s$ than that by other methods, especially when $p$ is large. So, compared with other methods, the source spreaders selected by VoteRank are more decentralized in the whole network. Actually, as pointed out in Ref. \cite{HuEPL2014}, the spreading will be more effective when $L_s$ gets larger.

\section*{\textcolor[rgb]{1.00,0.00,0.00}{Computational complexity analysis}}
\textcolor[rgb]{1.00,0.00,0.00}{The total computational time includes three parts as follows: the time of initializing voting ability and voting score, the time of selecting a node with the highest voting score, and the time of updating the voting ability and voting score. For the first part, the time of initializing voting ability is $O(n)$ and that of initializing voting score is $O(<k>n)=O(m)$ where $<k>$ is the average degree of network and $m$ is the number of edges, so, the computational complexity of this step is $O(n+m)=O(m)$. Particularly, the computational complexity is $O(1)$ if we set initial voting ability as 1. For the second part, the computational complexity is $O(n)$ so as to select a node with the highest voting score. And if we take high efficient data structure such as red-black tree, the computational complexity will decrease to $O(\log n)$. For the third part, just the information of nodes with a distance of 2 from the newly selected spreader needs updating. Hence, the computational complexity is $O(<k>^2)=O(m^2/n^2)$. To select $r$ spreaders with $r$ times in step 2 and 3, the total computational complexity is $O(m)+O(r \log n)+O(r<k>^2)=O(m+ r\log n+ r m^2/n^2)$. If networks is sparse and $r\ll n$, the computational complexity of VoteRank can be reaching $O(n)$. Although above analysis is based on undirected network, it is similar for case of directed network.}

\section*{Discussion}
In summary, with utilizing information of $r-1$ ranked nodes to rank the $r^{th}$ node, we get an  obvious boost on information spreading in complex networks, especial in large scale networks. However, when $r$ is small, little information can be utilized and the advantage of VoteRank is not significant. When $r$ becomes large, the information accumulated by the $r-1$ previous nodes becomes abundant and can make a significant improvement. VoteRank provides a simple yet effective way to determine the next most influential node based on the selected nodes. It is worth mentioning that VoteRank outperforms K-shellRank on undirected network, and also outperforms other benchmark algorithms such as PageRank, ClusterRank and IndegreeRank on directed network. The results also indicate that performance of VoteRank is \textcolor[rgb]{1.00,0.00,0.00}{fairly} stable with different infected rate $\lambda$ and different scale of initial spreaders $p$ in terms of information spreading.
\textcolor[rgb]{1.00,0.00,0.00}{ Another interesting question is how to judge the optimal number of $r$ to get the best spreading ability.  In fact, this problem has two equal forms. The first is maximizing the spreading ability while fixing the number of initial spreaders. The second is minimizing the number of initial spreaders for giving spreading ability, i.e., fixing the number of final affected nodes. We just take into account the first form in our work, the other form can be analyzed similarly. Besides, some researchers use the robustness $R$ \cite{Schneider2011PNAS}, which is defined as $R=\frac{1}{n}\sum_{i=1}^{n} \sigma(i/n)$  where $\sigma(i/n)$  is the fraction of nodes belonging to giant component after removing $i/n$ of nodes from network, to evaluate the attacking ability of a method. The method has higher attacking ability if $R$ is smaller. }
In recent years, many researchers aimed to the study of temporal networks, including structure and dynamics \cite{Peter2012PR}. To identify a set of influential nodes in temporal networks is an important and interesting topic. How to extend our work to temporal networks is worth further studying.

\section*{Author contributions}
J.-X. Z. and D.-B. C. designed the research and prepared all figures. J.-X. Z. and Z.-D. Z. performed the experiments.  D.-B.C. and Q. D. analyzed the data. All authors wrote and reviewed the manuscript.
\section*{Acknowledgments}
This work is partially supported by the National Natural Science Foundation of China under Grant No. 61433014,  by the Fundamental Research Funds for the Central Universities under Grant Nos. ZYGX2014Z002, ZYGX2015J152 and ZYGX2015J156 and by the Shanghai Research Institute of Publishing and Meida under Grant No. SAYB1402.
\section*{Additional information}
Competing financial interests: The authors declare no competing financial interests.
\end{document}